\documentclass[aps,prl,twocolumn]{revtex4}

\usepackage{graphicx}

\begin{document}
\DeclareGraphicsExtensions{.eps,.jpg,.pdf,.mps,.png}
\title{Enhancing the photomixing efficiency of optoelectronic devices in the terahertz regime}

\author{Subrahmanyam Pilla}
\affiliation{Department of Physics, University of California, San
Diego, CA 92093} \email{manyamp@gmail.com}

%\date{\today}
\begin{abstract}
A method to reduce the transit time of majority of carriers in
photomixers and photo detectors to $< 1$ ps is proposed. Enhanced
optical fields associated with surface plasmon polaritons, coupled
with velocity overshoot phenomenon results in net decrease of
transit time of carriers. As an example, model calculations
demonstrating $> 280\times$ (or $\sim$2800 and 31.8 $\mu$W at 1
and 5 THz respectively) improvement in THz power generation
efficiency of a photomixer based on Low Temperature grown GaAs are
presented. Due to minimal dependence on the carrier recombination
time, it is anticipated that the proposed method paves the way for
enhancing the speed and efficiency of photomixers and detectors
covering UV to far infrared communications wavelengths (300 to
1600 nm).
\end{abstract} \maketitle

Recent experimental observation of extraordinary optical
transmission through a normal metal film having subwavelength
holes\cite{Ebbesen} has generated intense interest in harnessing
the underlying physics for photonic applications. It is widely
believed that quasi two-dimensional electromagnetic excitations
tightly bound to the metal surface known as Surface Plasmon
Polaritons (SPPs) are responsible for the observed near field
enhancement of the optical radiation. Exponential decay of field
components arising from SPPs (penetrating $\sim$100, $\sim$10 nm
in dielectric and metal respectively), make them highly attractive
for miniature photonic circuits and devices\cite{Barnes}.
Currently, several potential applications of SPPs are being
explored including wave guiding\cite{Bozhevolnyi} and near-field
microscopy\cite{Furukawa}. However, application of SPPs to enhance
the speed and efficiency of photo detectors or photomixers has not
been attempted due to the complexity of underlying physics and
optimal device designs involving complicated dielectric-metal
structures.
\begin{figure}
\begin{center}
\includegraphics[width= .99\linewidth]{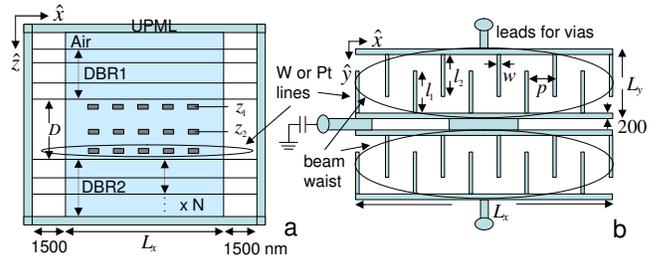}
\end{center}
\caption{Schematic of FP cavity (a) and a layer of interdigitated
metal lines (b) modelled using FDTD and FD simulators. The FP
cavity is formed by 3 or 4 dielectric layers constituting DBR$_1$,
$N$ pairs of low/high refractive index dielectric layers
constituting DBR$_2$, and absorbing layer of thickness $D$. The
active volume is terminated by 15 cell UPML layers in FDTD
implementation. The lines parallel to the $x$ axis (width 200 nm)
are connected to either signal transmission lines or planar
antenna fabricated on top of the absorbing layer (not shown).
Through layer vias electrically connect identical layers of
interdigitated lines (b) placed at various depths $z_i$ within the
absorbing layer. The overall dimensions of the absorbing volume
(excluding DBRs and via leads) are given by $L_x,~L_y,~D$ whereas
number of interdigitated metal layers, their vertical positions,
(measured from the Air-DBR$_1$ interface), the metal line (finger)
thickness, pitch, width, and lengths of +ve and -ve electrodes are
given by $i,~z_i,~d,~p,~w,~l_1,~l_2$ as shown. $x = 0,~y = 0$ is
located at the center of top half of the interdigitated pattern in
(b). When the top and bottom interdigitated structures in (b) are
connected in series as shown with the center electrode grounded,
the resulting structure can be excited by lasers forming two
quasi-elliptical spots. Such a radiation pattern is achievable by
choosing proper laser propagation mode (such as TEM$_{01}$ mode).
The spacing between the two halves is adjusted to match the beam
pattern of the specific laser system being used.}
\label{optical_column}
\end{figure}

In this letter we propose a simple interdigitated metal structure
embedded in a suitable semiconductor material to harness SPPs for
improving the speed and efficiency of photomixers and detectors.
Such a device can be realized with the existing material growth
techniques\cite{Harbison}. As an example, we demonstrate via model
calculations, significant ($> 280\times$) improvement in the
efficiency of a THz photomixer based on Low Temperature grown GaAs
(LT-GaAs). Microwave to THz radiation sources in the 0.1 to 10 THz
are being extensively studied for their application in
communications, medical imaging, and
spectroscopy\cite{Bjarnason,Mittleman,Pearson,Woodward,Kawase}.
For many applications- compact, narrow-linewidth, widely tunable
continuous wave sources are necessary. In particular, to be used
as a local oscillator in communications systems the THz source
should produce stable output power $> 10~\mu$W\cite{Pearson,Ito}.
Among the various techniques being
pursued\cite{Bjarnason,Pearson,Ito,Dohler,Kohler} electrical down
conversion of optical microwave sources in a suitable
photomixer\cite{Bjarnason,Ito,Dohler} are appealing due to the
easy availability of tunable, narrow-linewidth, and stable solid
state lasers. However, until now the output power from these
photomixers is limited to $< 10~\mu$W in the crucial THz range. In
a conventional Metal-Semiconductor-Metal (MSM)
photomixer\cite{Brown}, optically thick (~100 nm) interdigitated
metal lines are fabricated on a high resistivity semiconductor
with subpicosecond recombination time of carriers ($\tau_e/\tau_h
$ for electrons/holes respectively) and high breakdown field
limit. So far, only a few materials such as
LT-GaAs\cite{Bjarnason,Brown} and Fe implanted InGaAs\cite{Suzuki}
have been shown to meet these requirements. From the photomixer
theory, when two lasers (wavelengths $\lambda_1,\lambda_2$,
$\lambda_0 = (\lambda_1+\lambda_2)/2$, and powers $P_1 = P_2 =
P_0/2$) with their difference frequency $f =
c(\lambda_1^{-1}-\lambda_2^{-1})$ in the THz range are incident on
a dc biased photomixer\cite{Brown}, the THz wave output power
$P_f$ is given by
\begin{equation}\label{Eq1}
P_f = \frac{\frac{1}{2}R_L i_p^2}{[1+(2\pi f(R_L+R_S)C)^2]},
\end{equation} where $i_p$ is the photocurrent generated in the external circuit,
$c$ is velocity of light in free space, $C$ is the capacitance,
$R_S$ is small internal resistance of the metal structure, and
$R_L$ is the load resistance in the 72-200
range\cite{Bjarnason,Dohler,Brown}. To accelerate carriers
generated deep inside the semiconductor, high dc voltage ($V_b
\approx 40$ V) is applied across the electrodes. This results in
fields quickly exceeding the breakdown limit near the electrodes
leading to device failure. As the carrier transport is transit
time ($t_{tr}$) limited, i. e., $t_{tr} > f^{-1}$, recombination
time $\tau \ll f^{-1}$ is required to recombine the carriers that
do not contribute to $i_p$ before the beat cycle
reverses\cite{Dohler,Brown}. Even if the photomixer is placed in a
suitable optical cavity, due to strong reflection from thick metal
electrodes, no carriers are generated directly below the
electrodes where $t_{tr}$ will be small. In addition,
subwavelength features of the metal lines produce strong near
field diffraction patterns that are not taken into account in
conventional designs\cite{Bjarnason,Dohler,Brown}.

\begin{figure}
\begin{center}
\includegraphics[width= .9\linewidth]{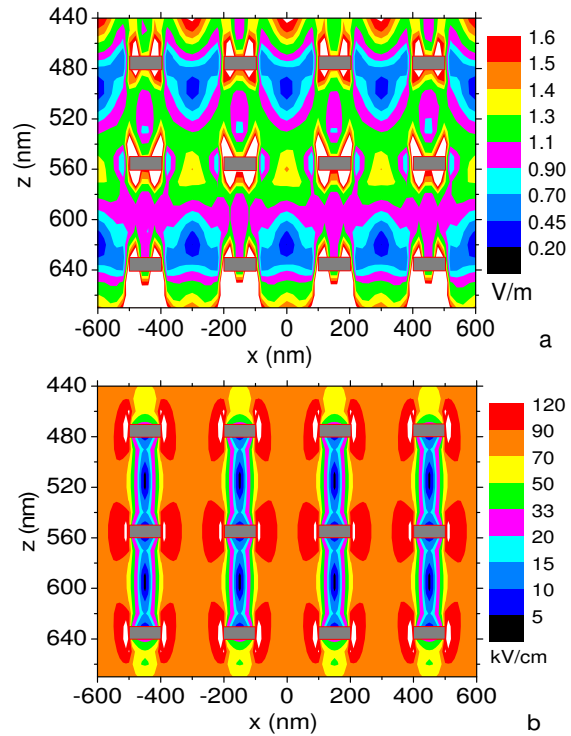}
\end{center}
\caption{Optical electric field amplitude (a) and dc electric
field strength (b) inside the $D = 230$ nm LT-GaAs absorbing layer
of the photomixer design of Fig 1 optimized for $\lambda_0 = 850$
nm. Incident optical field amplitude $E_x^0 = 1$ V/m and $V_b =
\pm1$ V. For clarity, fields in only the $x = \pm600$ nm region
along the $y = 0$ plane are plotted. The grey rectangular regions
in the plots show the cross-section of the interdigitated W lines
with $L_x = 7000,~L_y = 2500,~d = 10,~ z_1 = 475,~ z_2 = 555,~ z_3
= 635,~p = 300,~w = 100,~l_1 = 1500,~l_2 = 1400$ nm respectively.
$C = 4.90$ fF and $R_S = 2.4~\Omega$ for this device. The FP
cavity is formed by three layers of (TiO$_2$, Si$_3$N$_x$,
CaF$_2$) constituting DBR$_1$ (refractive indices (2.4, 2.0, 1.23
to 1.36)\cite{Heavens} and thicknesses (100, 150, 190) nm for each
layer respectively), four pairs of (Al$_2$O$_x$, GaAs)
layers\cite{Park} forming DBR$_2$ (refractive indices (1.6,
$3.53-0.068i$) and thicknesses (130, 60) nm), and $D \approx
\lambda$ absorbing LT-GaAs layer with refractive index
$3.77-0.068i$.} \label{optical_column}
\end{figure}

In the present technique, we exploit SPPs for generating more
carriers close to subwavelength normal metal electrodes embedded
in a photoconductor layer sandwiched between two Distributed Bragg
Reflectors (DBR) that form a Fabry-Perot (FP) cavity. When coupled
with velocity overshoot phenomenon reported in many photoconductor
materials\cite{Dohler,Betz,Awano,Reklaitis,Thobel,Foutz}, $t_{tr}$
of majority of carriers is reduced to $<$ 1 ps, i.e., $t_{tr} \leq
f^{-1}$. For efficient use of SPPs, it is desirable to have thin
($\sim$10 nm) normal metal lines with subwavelength features
distributed throughout the active volume (Fig 1a) in a manner that
would collectively enhance the optical field intensity in the
vicinity of the metal lines. This must however be accomplished
without significantly increasing $C$ and $R_S$. We show that for
the proposed structure (Fig 1) optimized for $\lambda_0 =$ 850 nm,
$P_f$ is $\sim$2800 and 31.8 $\mu$W at $f = 1$ and 5 THz
respectively. A convenient way to model such a complex structure
is to solve the Maxwell's equations with appropriate boundary
conditions using Finite Difference Time-Domain (FDTD)
formulation\cite{Yee,Taflove,Sadiku}. A 3D-FDTD simulator with
Uniaxial Perfectly Matched Layer (UPML)\cite{Taflove} surrounding
the photomixer is developed for this purpose. Along the z-axis,
the active volume consists of a few hundred nm of free space
followed by DBR$_1$, absorbing region with complex structure of
normal metal electrodes, and pairs of low/high refractive index
layers constituting DBR$_2$ (Fig 1a)\cite{Park}. For a given $D$,
the refractive index and thickness of DBR layers are first
optimized by calculating the reflection and transmission
coefficients of the FP cavity using matrix methods\cite{Heavens}.
The FP cavity is excited by a linearly polarized, plane wave
propagating along $+z$ direction, with gaussian (elliptic)
intensity profile in the $x-y$ plane. The source plane is placed
in the free space above the DBR$_1$. Frequency-domain Lorentz
dispersion model\cite{Judkins} is adapted for metals with negative
dielectric constant such as W or Pt.

Figure 2a shows the FDTD results of a three layer stack of
interdigitated W lines embedded in  230 nm LT-GaAs absorbing layer
of a photomixer design of Fig 1 optimized for $\lambda_0 = 850$
nm. The plot clearly shows the near field enhancement resulting
from the thin normal metal electrodes. In the absence of the W
electrodes, we obtain electric field amplitude maxima (maximum
amplitude ~1.4 V/m) corresponding to three antinodes in the
standing wave formed between the DBRs. The 3D internal dc electric
fields and $C$ are computed by solving Laplace's equation with
appropriate boundary conditions using FD techniques\cite{Sadiku}.
Figure 2b shows the contour plot of the static electric field
strength within the LT-GaAs layer. The data show that unlike the
traditional MSM structure, the field strength in this design is
well above the critical field ($\sim$5 kV/cm). Moreover, the field
strength is $\sim$90 kV/cm between neighboring electrodes, in
particular at the center of the device where most of the carriers
are generated due to gaussian intensity profile of the incident
laser beams. In the rest of the volume it has a broad peak at 18
kV/cm (not shown). Results show that $C = 4.90$ fF and $R_S =
2.4~\Omega$  based on the resistivity of thin annealed or
epitaxial W films ($\sim$$5~\mu\Omega-$cm)\cite{Elbaum}. The
highest electric field inside the device is about four times lower
than the breakdown field (500 kV/cm); therefore device failure due
to electric breakdown is unlikely for this $V_b = \pm1$ V.
\begin{figure}
\begin{center}
\includegraphics[width= .75\linewidth]{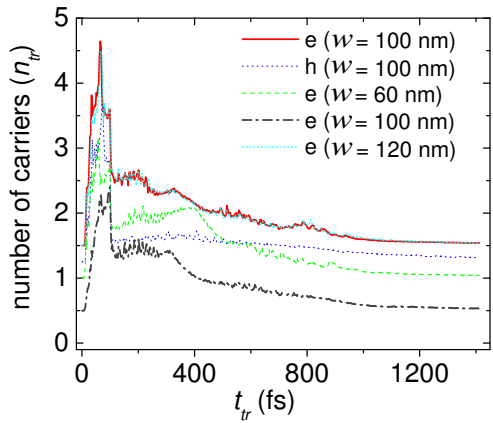}
\end{center}
\caption{Carrier transit time distribution $n_{tr}(t_{tr})$ for
electrons (e) and holes (h) in the photomixer design of Fig 1. The
parameters for red (solid) and blue (dot) curves are same as in
Fig 2 with $w = 100$ nm where as for grey (dash-dot) curve only
polarization is changed to $E_y$. For the green (dash) and cyan
(short dash) curves ($L_x,~p,~w$) are changed to (7560, 300, 60)
and (7480, 320, 120) nm respectively while rest of the parameters
remaining unaltered from those of Fig 2. The photomixer is excited
by an 850 nm source at $P_0 = 60$ mW for all structures. For
clarity, the curves are shifted along $y-$axis and the relative
shift can be easily obtained from the values at $t_{tr} = 0$ fs.}
\label{optical_column}
\end{figure}

Photocurrent $i_p$ is calculated by first computing the
electron/hole transit time distributions $n_{tr}^e(t_{tr}),~
n_{tr}^h(t_{tr})$ shown in Fig 3 for the entire absorbing volume
based on the above FDTD and FD calculations. From the data
available in literature for GaAs and LT-GaAs, for $t_{tr} \leq
t_{onset} =$ 100 fs, the carrier motion can be approximated by
ballistic transport with electron effective mass $m_e = 0.088
m_0$, where $m_0$ is electron rest mass. This value is consistent
with the slope of the linear portion of transient drift velocity
curve obtained from Monte-Carlo calculations\cite{Reklaitis}. The
corresponding effective mass for holes is $m_h = 0.106 m_0$.
Effective masses larger than the accepted values for GaAs (0.063,
0.076 $m_0$ for electron and light holes respectively) are
considered so that longer $t_{tr}$, thereby lower estimate of
$i_p$ results from these calculations. In the photomixer of Fig 2,
pure ballistic transport is applicable to electrons generated
close to the +ve electrodes and holes generated close to -ve
electrodes. These carriers generated predominantly in the near
field enhancement region, transit through non-uniform dc fields in
the 5 to 90 kV/cm range. It should be noted that $t_{onset} =$ 100
fs is considerably lower than the theoretical limit of ballistic
motion in GaAs for this field strength range\cite{Betz}. For
$t_{tr} \geq t_{onset}$, electron motion is approximated by
qasiballistic transport with time dependent velocity distribution
similar to ref. 18 up to $t_{tr} =~3$ ps, and equilibrium drift
velocity ($\sim1 \times 10^7$ cm/s) for $t_{tr} >$ 3 ps. Hole
motion is approximated by one third of the electron velocity at
any given $t_{tr} \geq t_{onset}$ resulting in velocities lower
than those reported for GaAs\cite{Awano}.

\begin{figure}
\begin{center}
\includegraphics[width= .8\linewidth]{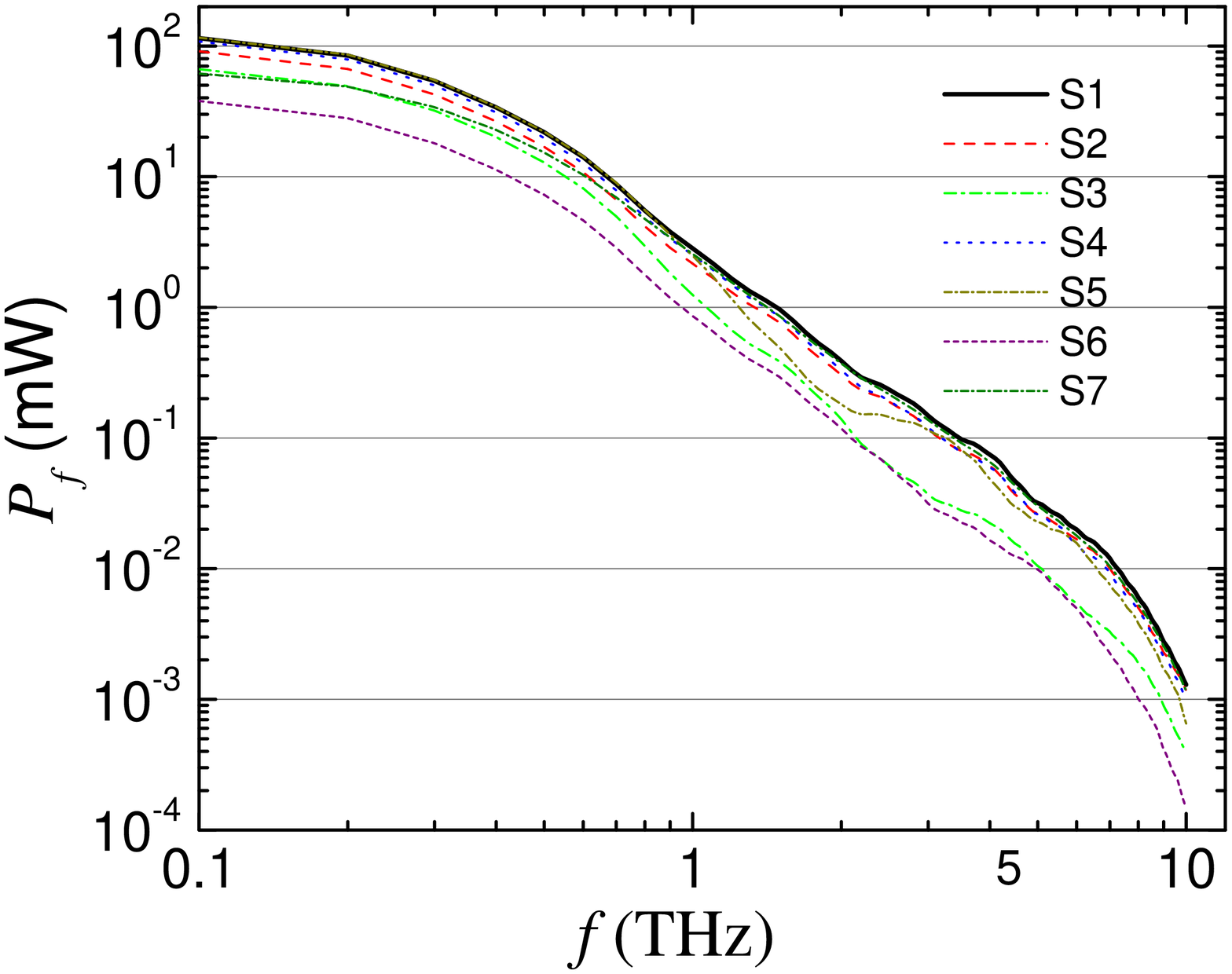}
\end{center}
\caption{Calculated THz output power $P_f$ from the photomixer
design of Fig 1 excited by $\lambda_0 = 850$ nm radiation with
incident power $P_0 = 60$ mW. $\tau = 4$ ps for designs S1 to S6
where as for S7 $\tau = 1$ ps. Parameters for S1 are same as in
Fig 2 whereas for S2 $z_3 = 645$ nm, for S3  $z_3 = 645$ nm and
incident radiation is $E_y$ polarized, for S4 ($p,~w$) are (320,
120) nm respectively, for S5 ($L_x,~w$) are (7560, 60) nm
respectively, and for S6 ($z_1,~z_3$) are (455, 655) nm
respectively. The remaining unlisted parameters are the same in
all cases. In all designs the interdigitated lines in different
layers have no relative orientation angle differences.}
\label{optical_column}
\end{figure}

The integral of curves in Fig 3 give the carriers ($N_e,~N_h$ for
electrons and holes respectively) generated per period
($T\simeq2.83$ fs) of the 850 nm source with $P_0 = 60$ mW. For
the parameters of Fig 2, $\sim$98\% of incident power is absorbed
in LT-GaAs layer while $n_{tr}$ has peaks at $t_{tr}\approx$ 65
and 70 fs for electrons and holes respectively (solid red and
dotted blue curves). Such a sharp peak followed by several
satellite peaks and a long tail of the $n_{tr}$ distribution can
be understood from the fact that most of the carriers are
generated close to the electrodes in the near field enhancement
regions where the static fields are also strong. The satellite
peaks result from periodicity of the electrode structure and
specific choice of the velocity distribution. As $t_{tr} <$ 1 ps
for majority of carriers in Fig 3, $\tau$ is not a critical factor
in determining the performance at THz frequencies. A sharp drop in
$n_{tr}$ by a factor of 2 for electrons and a factor of 3 for
holes at $t_{tr}\approx t_{onset}$ is due to the lower estimate of
carrier velocities for $t_{tr} > t_{onset}$ resulting from the
uniform field qasiballistic distribution function\cite{Reklaitis}
applied to a case where fields are inherently inhomogeneous.
Although, the above choice of velocity distribution for
quasiballistic motion is consistent with the fact that over a
large volume fraction of the absorbing layer the static field is
$\sim$18 kV/cm (not shown), a rigorous calculation should include
quasiballistic velocity distribution appropriate for inhomogeneous
fields. This would require ensemble Monte Carlo and carrier
trajectory calculations carried out simultaneously. However, above
calculations are adequate for obtaining the lower limit of the THz
power output because a rigorous calculation would probably produce
a more uniform $n_{tr}(t_{tr})$ distribution around $t_{tr}\approx
t_{onset}$ without altering the distribution for $t_{tr} <
t_{onset}$ or the integral of $n_{tr}(t_{tr})$. This will further
reduce the number of carriers in the long tail of the $n_{tr}$
distribution.

Based on above $n_{tr}(t_{tr})$ distributions, the number of
electrons captured in the metal electrodes at time $t$ will be
\begin{equation}\label{Eq2} n_{cap}^e(t) = \int\limits_{t_c =
0}^{t_c=t}n_{gen}^e(t_c)n_{tr}^e(t-t_c)e^{-(t-t_c)/\tau_e}~dt_c,
\end{equation} where $n_{gen}^e(t) = \frac{N_ec}{\lambda_0}[1+\cos(2\pi f t]$ is
the number of electrons generated in LT-GaAs layer per second and
$t_c$ is the carrier creation time. Similar expressions can be
written for number of holes captured $n_{cap}^h(t)$ and the
electrons/holes available for conduction in the photomixer
$n_{avl}^e(t),~n_{avl}^h(t)$. In the calculation of
$n_{cap},~n_{avl},~i_p(t) = e(n_{cap}^e(t)+n_{cap}^h(t))$ ($e$ is
electron charge), and $P_f$, we set $R_L = 72~\Omega,~\tau = 2
\tau_e = 2 \tau_h/3$\cite{Brown}, and varied $\tau$ in the 0.5 to
6 ps range. For the parameters of Fig 2, the steady state electron
density $\overline{n}$ obtained from the dc value of
$n_{avl}^e(t)$ for $\tau =~4$ ps is $\sim6 \times 10^{15}$
cm$^{-3}$ (at $P_0 =$ 60 mW) with holes 1.5 times more numerous
than electrons. We estimated that dipole fields arising from this
space charge are negligible ($<$ 2.5\%) in comparison to strong
fields (Fig 2b) present in the photomixer.

We have carried out calculations for over 70 configurations by
systematically varying $D,~L_x,~L_y,~d,~z_i,~p,~w,~l_1,~l_2,~i$,
polarization of the lasers ($E_x$ or $E_y$), orientation of
interdigitated lines in different layers, and DBR parameters. Fig
4 shows the $P_f$ values obtained from Eq 1 and 2 for some of the
configurations in the 0.1 to 10 THz range. The data shows that
$P_f \propto f^{-2.77}$ in the 0.5 to 6.5 THz range in contrast to
$f^{-4}$ roll-off of $P_f$ for a conventional
photomixer\cite{Brown}. A recent nip-nip photomixer concept is
shown to have $f^{-2}$ roll-off for $f < $1.5 THz\cite{Dohler}.
Therefore, the design of Fig 1 exploiting SPPs offers significant
improvement over the existing photomixer designs. Based on 3D FD
computation of steady-state heat equation with appropriate thermal
boundary conditions for the device parameters of Fig 2 ($P_0 = 60$
mW, $V_b = \pm1$ V), internal temperature ($T_i$) of the device is
estimated to be $\sim$200 K above the substrate temperature
requiring substantial cooling (to 77 K) when $P_0$ is high.
Thermal conductivity ($k$) of various DBR layers and that of
LT-GaAs are approximated by assigning $k_\parallel^{\rm{LT-GaAs}}
= $ 46 Wm$^{-1}$K$^{-1}$ and $k_\perp^{\rm{LT-GaAs}} =
k_\parallel^{\rm{DBR}} = k_\perp^{\rm{DBR}} =$ 10
Wm$^{-1}$K$^{-1}$, where $k_\parallel/k_\perp$ are
in-plane/out-of-plane conductivities. This internal heating
therefore limits $P_0$ to $\sim$60 mW. However, the structure of
Fig 1b, which is equivalent to two capacitors in series coupled to
the antenna or transmission line, offers an alternative. If it is
excited by TEM$_{01}$ mode lasers with total power $P_0 =$ 120 mW
as shown, the output can be increased to 2$P_f$ (of Fig 4) without
worsening $T_i$ or $\overline{n}$. To demonstrate the
applicability of this method to other semiconductor materials,
similar calculations are carried out for two more photomixers
based on Be doped In$_{1-x}$Ga$_x$As and GaN optimized for
operation at $\lambda_0 =$ 1550 and 343 nm respectively. Both
structures show strong near field enhancement arising from SPPs
similar to Fig 2a and further work is underway to calculate the
device efficiencies. Author wishes to thank John Goodkind for
introducing him to the fascinating subject of Auston switches.

%\begin{figure}
%\begin{center}
%\includegraphics[width= .65\linewidth]{}
%\end{center}

%\caption{}
%\label{}
%\end{figure}

\end{document}